# Virtual Breakdown Mechanism: Field-Driven Splitting of Pure Water for Hydrogen Production

*Yifei Wang[1], S. R. Narayanan[2] and Wei Wu[1]**

*wu.w@usc.edu

1. Ming Hsieh Department of Electrical Engineering, University of Southern California
2. Department of Chemistry, University of Southern California

***Abstract***- Due to the low conductivity of pure water, using an electrolyte is common for achieving efficient water electrolysis. In this paper, we have broken through this common sense by using deep-sub-Debye-length nanogap electrochemical cells for the electrolysis of pure water. At such nanometer scale, the field-driven pure water splitting exhibits a completely different mechanism from the macrosystem. We have named this process "virtual breakdown mechanism" that results in a series of fundamental changes and more than $10^5$-fold enhancement of the equivalent conductivity of pure water. This fundamental discovery has been theoretically discussed in this paper and experimentally demonstrated in a group of electrochemical cells with nanogaps between two electrodes down to 37 nm. Based on our nanogap electrochemical cells, the electrolysis current from pure water is comparable to or even larger than the current from 1 mol/L sodium hydroxide solution, indicating the high-efficiency of pure water splitting as a potential for on-demand hydrogen production.

As a clean and renewable resource to substitute fossil fuels in future, efficient hydrogen production has become increasingly important. Moreover, hydrogen has huge demands in a variety of industrial fields such as metal refining, ammonia synthesis, petroleum refining and energy storage. Today, more than 90% of the industrially-produced hydrogen comes from steam reforming of natural gas and gasification of coal and petroleum coke[1]. However, these hydrocarbon resources are non-renewable and generate greenhouse gases as by-products. Water electrolysis, especially when connected to renewable energy supplies such as wind turbines, solar photovoltaic, and hydroelectric generation, can provide high-purity hydrogen and could be a solution for a sustainable energy supply. However, at present only 4% of the industrial hydrogen production comes from water electrolysis[2-4], basically due to the low conversion efficiency resulting from the high cell voltage, which arises from the large overpotential at the electrodes and ohmic loss in the solution, especially at large operating current density[5]. Photolysis of water is another promising technology[6-8]. However, this method is still under development because of low quantum efficiencies.

Water electrolysis has been known for more than 200 years[9] and applied on industrial hydrogen production for over 100 years[2, 10]. However, efforts to increase the energy efficiency and reduce the cost of water electrolysis continue even today. Research has focused on electrocatalytic materials[11-13], temperature and pressure effects[5, 14, 15], and optimization of electrolyzer design[16, 17]. Different from all of these foregoing approaches, we have demonstrated a new approach to improve the electrochemical reaction efficiency, by using electrochemical cells with distance between anode and cathode in nanometer-scale. With these nanogap electrochemical cells (NECs), pure water (without any added electrolyte) can be electrochemically split into hydrogen and oxygen efficiently, in contrary to the traditional thinking that pure water cannot be electrolyzed. Our experiments have demonstrated that the equivalent conductivity of pure water has been enhanced more than $10^5$-fold, and the performance of NECs with pure water can be comparable to or even better than with 1 mol/L sodium hydroxide solution, which results from a completely different microscopic mechanism of field-driven ions transport to enhance water ionization and even virtual breakdown. Compared to current industrial water electrolysis operated at 70-90 °C[2, 3] with strong alkaline electrolyte, our NEC design with pure water can eliminate difficulties of working with strong alkaline electrolyte and also avoid the need for high temperatures, showing a great potential for high energy-efficiency on-demand hydrogen production for both mass manufacturing and portable devices.

For simplicity, consider the solution resistance between anode and cathode for water splitting, as given by

$$R = \rho \frac{l}{S} \tag{1}$$

where $\rho$ is the resistivity, $l$ is the resistor length (electrode distance) and $S$ is the cross-section area of the resistor. We found that as the electrode distance shrinks to much smaller than Debye-length $\lambda_d$ (around 1 μm for pure water), not only the value of $l$ decreases, the equivalent resistivity $\rho$ decreases greatly as well, which in fact contributes more to the decrease of resistance $R$. This is attributed to the huge electric field between two electrodes within such deep-sub-Debye-length region (Figure 1). For water electrolysis with strong electrolyte in macrosystem, the electric field is screened by the double layer, leading to nearly zero electric field in bulk solution (Figure 1(a)). However, when the electrode gap distance is smaller than the Debye-length, large electric field can be uniformly distributed in the entire gap due to overlapping of the double layers at the two electrodes. In our metal-dielectric-metal sandwiched-like NECs, the gap distance is tuned by adjusting the silicon nitride thickness and can be easily achieved to deep-sub-Debye-length in pure water. Figure 1(b) shows the simulation results of electric field distribution between two electrodes with different gap distances (see details in Supplementary information). Close to the electrode regions both the nanogap cell and the macrosystem present a high electric field due to the double layer; however, in bulk solution the electric field in 100 μm macrosystem is only 10 V/m while in 0.1 μm gap the field can obtain $10^7$ V/m. Such a high electric field in the entire gap of nanogap cells can result in significant ion enrichment and ion migration[18, 19], and even further water ionization and virtual breakdown.

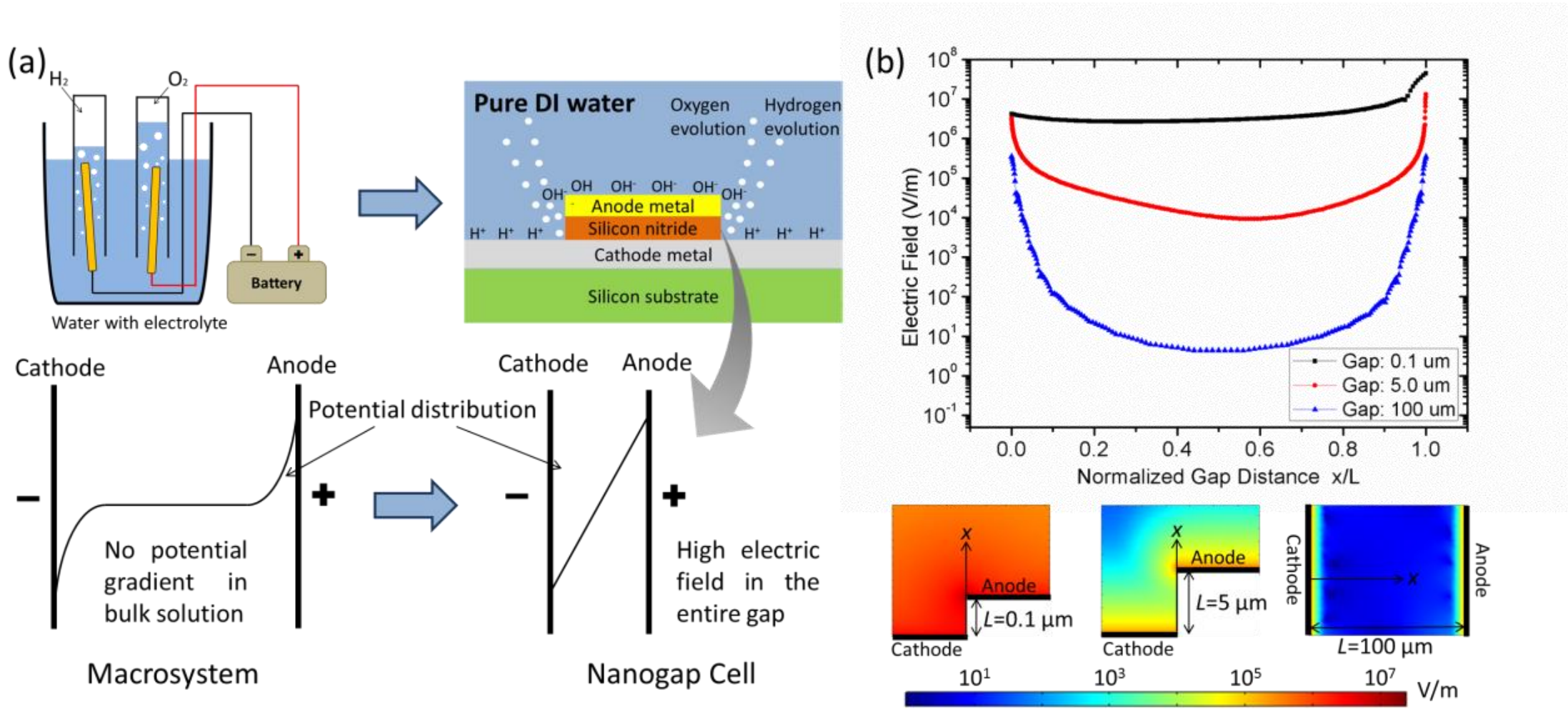

Figure 1. High electric field distributed in the entire gap between anode and cathode in NECs. (a) Schematic diagram of potential distribution comparison between macrosystem and our sandwiched-like nanogap cells. (b) Simulation results to show the electric field distribution (1-D plot and 2-D plot) between two electrodes with gap distance of 0.1 μm ($0.1\lambda_d$, sandwiched-like NEC), 5.0 μm ($5\lambda_d$, sandwiched-like NEC) and 100 μm (macrosystem, plate electrodes).

## Theoretical analysis

Figure 2(a) explains why pure water cannot be split efficiently in macrosystem, in which we take cathode and $H_3O^+$ ions as an example. Initially near the cathode surface water molecules can be dissociated into $H_3O^+$ and $OH^-$ ions. $H_3O^+$ ions obtain electrons from cathode leading to hydrogen evolution, while the newly-generated $OH^-$ ions can be transported very slowly through the bulk solution by slow diffusion or hopping process facilitated by a weak electric field in bulk solution. Moreover, the intrinsic concentration of $H_3O^+$ ions in bulk solution of pure water is too low to neutralize the $OH^-$ ions produced near the cathode. These lead to local $OH^-$ ions accumulation (so that the solution near cathode turns alkaline) especially at the cathode surface, causing the potential at the Helmholtz plane of the cathode to decrease (because of negatively-charged $OH^-$ ions). Such a potential decrease reduces the potential difference between the cathode and Helmholtz plane, further reducing the reaction rate of hydrogen evolution and thus water splitting. In other words, the reaction becomes very slow or even self-limited, showing a large equivalent resistance between cathode and anode. These phenomena also explain the rise in cathode overpotential, since a more negative cathode potential is necessary to allow the reaction to continue. The fundamental reason is the lack of rapid ions transport inside bulk solution.

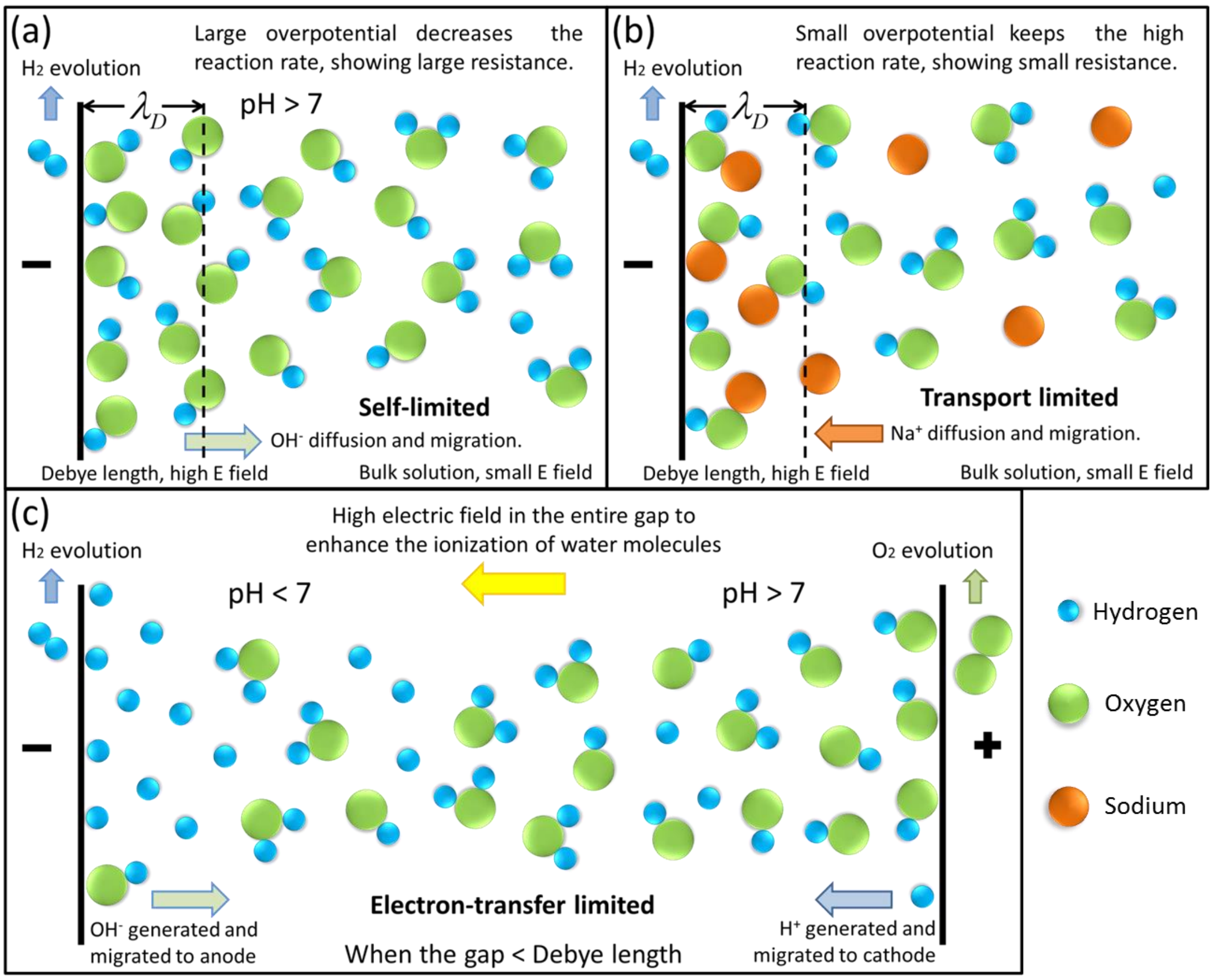

Figure 2. Schematic diagram of water splitting reactions in 3 different systems. (a) Pure water electrolysis in macrosystem is self-limited due to the lack of rapid ion transport inside bulk solution. (b) In sodium hydroxide solution, water splitting reaction can keep occurring but is limited by mass transport (mainly diffusion). (c) In nanogap cell, high electric field in the entire gap can enhance water ionization and mass transport (mainly migration), leading to efficient pure water splitting limited by electron-transfer, and completely opposite pH-value distribution compared to that in macrosystem.

When a high-concentration of sodium hydroxide is present in the electrolyte (Figure 2(b)), plenty of $Na^+$ ions from bulk solution can move to partially compensate for the charge from the newly-generated $OH^-$ ions near the cathode, restoring the potential difference between the cathode and Helmholtz plane, to reduce the overpotential requirement and sustain the reaction current. A similar process occurs at the anode. In this way, water electrolysis with strong electrolyte shows a small resistance between two electrodes. However, notice that even though the ions transport inside bulk electrolyte solution is large enough for the continued reaction, at cathode the sodium ions transport is still limited mainly by diffusion (because of weak electric field in bulk solution)[18, 20], which is often slower than $OH^-$ ions generation (*i.e.*, reaction electron-transfer). Under steady-state conditions, net $OH^-$ ions accumulation still occurs at cathode and the potential effect on Helmholtz plane still exists.

In pure water, when the counter-electrode is placed within the Debye-length (Figure 2(c)), the double layers of cathode and anode are overlapping with each other. Still at cathode, newly-generated $OH^-$ ions can be migrated rapidly from cathode towards anode due to large electric field in the entire gap. When the gap distance is small enough, initially the transport rate can be even higher than the electron-transfer rate. Once $OH^-$ ions are generated, they are immediately drawn from cathode to anode, resulting in the $OH^-$ ions waiting for electron-transfer at the anode, rather than accumulated at the cathode. The whole reaction would continue even in pure water, but now is limited by electron-transfer. In this case, net $OH^-$ ions accumulate near the anode and net $H_3O^+$ ions accumulate near the cathode, leading to completely opposite pH-value distribution compared to macrosystem (which maybe be good for protecting the anode against corrosion). Moreover, net $OH^-$ ion enrichment near the anode not only enhances the local reaction ions concentration but also increases the potential difference between anode and anode Helmholtz plane (which in fact decreases the overpotential requirement, as in the Frumkin effect[21]). According to Butler–Volmer equation[22],

$$j = Fk^0\left[C_O e^{-\alpha F(E-E^{0'})/RT} - C_R e^{(1-\alpha)F(E-E^{0'})/RT}\right] \quad (2)$$

such $OH^-$ ions accumulation can significantly increase the electrolysis current, namely water splitting throughput.

Under steady state, the field-driven effect is equivalent to the scenario that water molecules are split into $H_3O^+$ and $OH^-$ ions in the middle of the gap (see Figure S2), allowing $H_3O^+$ ions to drift towards the cathode and $OH^-$ ions to drift towards the anode, respectively. In other words, such huge electric field not only increases the transport rate, but also enhances the water molecules ionization (for RC-circuit model, see Figure S3). From a microscopic perspective, the conductivity of water has been enhanced "equivalently". From the equation of conductivity,

$$\sigma = nq\mu \quad (3)$$

where $q$ is the ion charge, $\mu$ is the ion mobility and $n$ is the ion concentration. Here the ion charges have not changed. The increased ion concentration only partially contributes to the conductivity. The fundamental change is that two half-reactions are coupled together, and the electric field distribution within the NEC gap leads to a significantly enhanced "apparent mobility". (In macrosystem, the intrinsic mobility cannot serve to the conductivity due to weak electric field in bulk solution.) The total effect looks like breakdown of pure water. However, notice that this effect is not traditional breakdown of pure water, which actually requires the electric field around 1 V/Å[23], about two magnitude orders larger than what we have discussed here. The high electric field in our NECs could not split water molecules directly; however, it enhances water ionization and ion transport, and thus equivalent pure water conductivity. That is why we called this field-driven effect, "virtual breakdown mechanism". The traditional view should be revised that even pure water can be conductive, when the electrode gap is small enough. This "virtual breakdown mechanism" can be applied on almost all types of weakly-ionized materials: such weak ionization actually helps to achieve the virtual breakdown effect.

## Device fabrication

There have been many efforts[24, 25] to fabricate nanogap electrodes. Electron/ion-beam lithographically-defined nanogap electrodes may not be scalable to large-area fabrication. Chemically-synthesized electrodes[26, 27] and mechanically-fabricated electrodes[28, 29] usually suffer from the lack of controllability. Sacrificial-layer based nanogaps[30-32] require complicated processes and thus perform poor yield[20] especially when nanogaps less than 100 nm. Bohn *et al*[18, 33, 34] and White *et al*[19, 20] have done excellent work on nanogap-based reversible redox cycling analysis at low ionic strength; however, their structures may not be suitable for irreversible reactions, especially with gas evolution.

The fabrication procedure of our open-cell sandwiched-like NECs is shown in Figure 3. First, a film stack of silicon dioxide (thermal oxidation), Pt (bottom cathode metal, electron-beam evaporation), silicon nitride (plasma-enhanced chemical vapor deposition) were deposited on silicon wafers. Afterwards, Ti (adhesion layer) with gold (top anode metal) and Cr (etching mask) were patterned by photolithography, electron-beam evaporation and lift-off process. The patterns consist of contact pads and 1-D gratings with different values of pitch. Here only the top gold anode was patterned and the bottom Pt cathode was a blank film. Next, the silicon nitride was etched with Cr (low sputtering yield[35]) as mask by low DC-bias anisotropic etching that was developed by us[36] (see Supplementary information), to avoid metal atoms sputtered out everywhere during etching. This method can avoid short-circuit between the top and bottom electrodes and thus enhance the yield of device fabrication. Finally, Cr mask was removed by Cr wet etching (ALDRICH®), which can also increase the hydrophilicity of the entire surface. The whole process is yield-controlled and can be scalable to mass manufacturing.

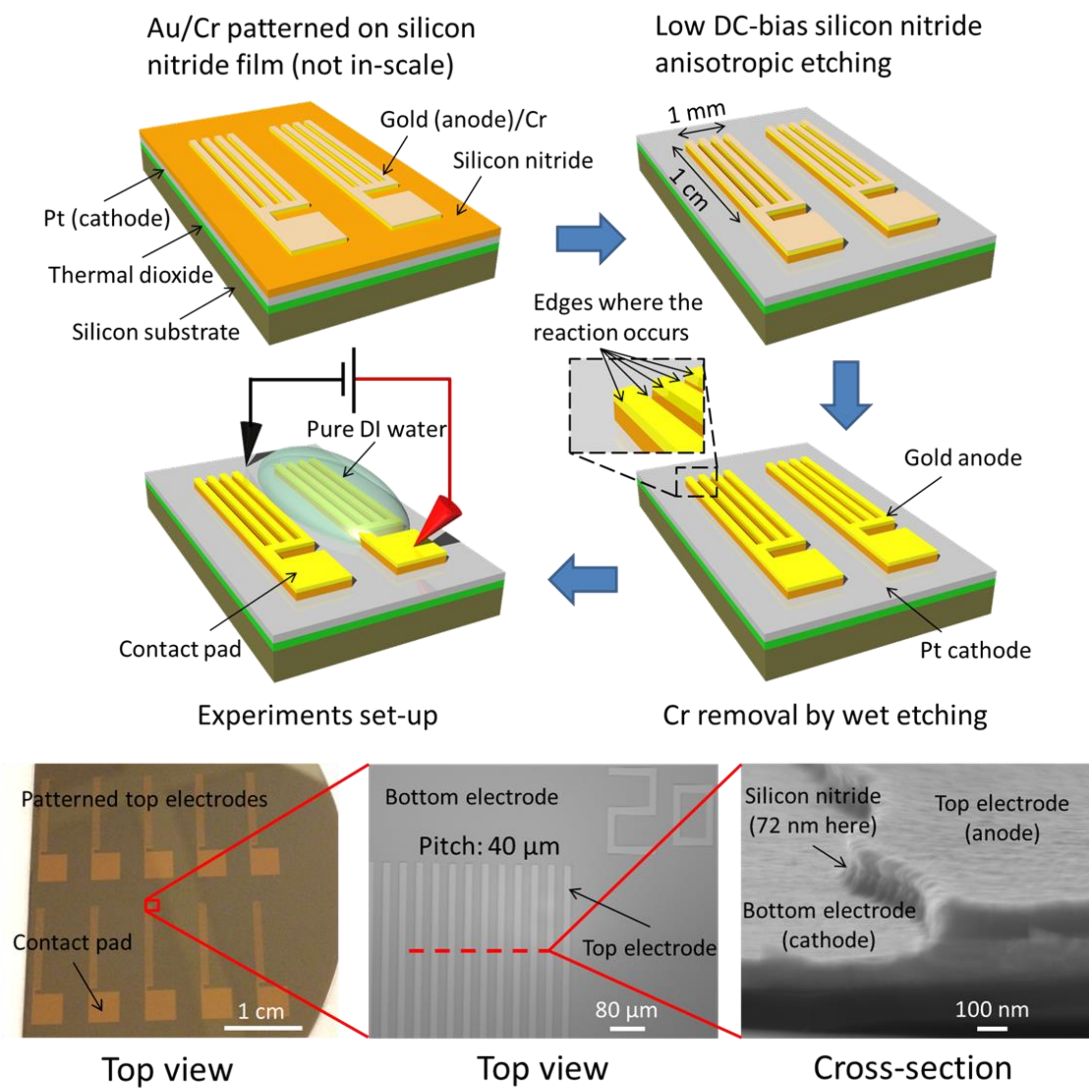

Figure 3. Fabrication procedures and results of our metal-dielectric-metal sandwiched-like NECs. This fabrication method can be simply applied on large area with high yield. Dimensions: the gap distance between the two electrodes, or thickness of silicon nitride, varied from 37 nm to 1.4 μm; thermal silicon dioxide, 100 nm thick; Pt, 100 nm thick; Ti, 2 nm thick; gold, 40 nm thick; Cr, 10 nm thick; the contact pads were 3.5 mm by 3.5 mm; the grating regions were 1 cm by 1 mm, with different grating pitches from 10 μm to 80 μm.

Platinum and gold were selected as the cathode and anode, respectively, due to their ability to catalyze hydrogen/oxygen evolution; gold is stable towards anodic oxidation[37] to avoid short-circuit between the two electrodes during electrolysis caused by metal dissolution and re-deposition[38]. The experimental set-up is schematically shown in Figure 3, with two electrode tips connected to the anode and cathode, and pure deionized water (DI water) was dropped to cover the grating region. The hydrophilicity of the entire surface guaranteed that the water completely wetted the whole electrode structure and gaps. Notice that the field-driven pure water splitting only occurs at the boundary (edges) of each grating line (more details in next section). Figure 3 also shows the fabrication results (40 μm grating pitch and 72 nm gap distance as an example) observed by unaided eyes (top view), by optical microscopy (top view) and by scanning electron microscopy (SEM) (cross-section view).

## Experiment results

**Pure water.** When exposed to air, $CO_2$ dissolution into water (pH around 5.7[39]) results in the Debye-length reduction from 1 μm to around 220 nm. For our smallest gap distance 37 nm, the double layer at each electrode has been at least compressed into 1/10 of the original Debye-length. At such deep-sub-Debye-length range, the uniform electric field in the entire gap is inversely proportional to the gap distance at a given voltage. Figure 4 shows the I-V curves from pure water experiments based on different gap distances from 37 nm to 1.4 μm. When gap distance shrank, the electrolysis current became larger due to higher electric field between two electrodes (Figure 4(a)). A voltage plateau around 0.9 V was observed in the log current *vs.* voltage plot (Figure 4(b)), which was independent of the gap distance. This may result from the dissolved oxygen reduction since the DI water was not saturated with inert gas; another reason may be surface oxide formation on gold during water electrolysis[38, 40, 41]. The entire surface became more hydrophilic after first test, which was consistent with the oxidation formation or hydroxide bond residue. In experiments, sometimes anode damage occurred when voltage was above 5 V (see Figure S5). Figure 4(d) shows part of the experimental set-up and bubbles generation around 2 V during the pure water splitting (see Figure S6 for more bubble effects).

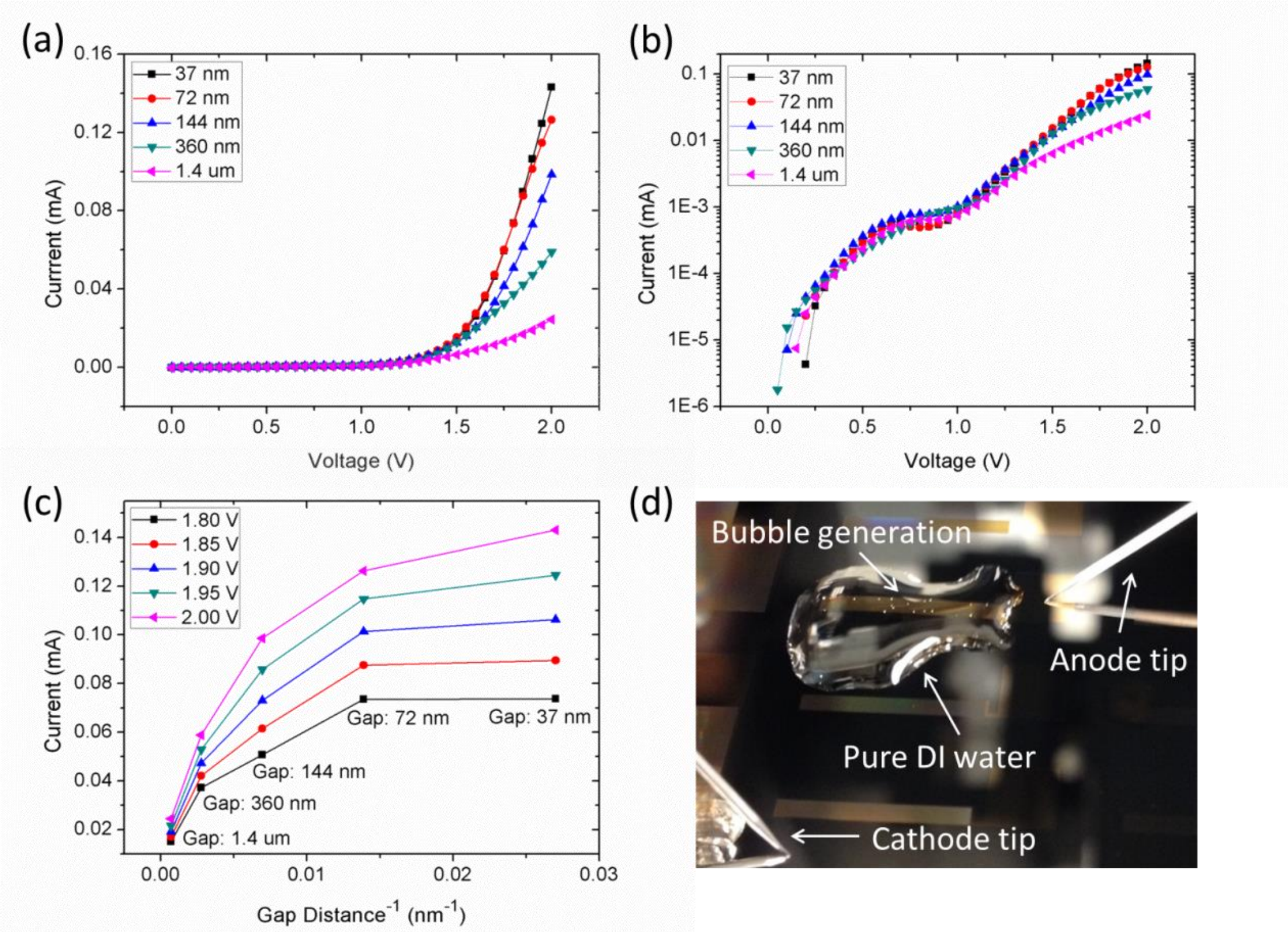

Figure 4. I-V curve measurements based on our NECs with pure DI water. The experiment conditions were 22 °C, 1 atm, humidity: 45%, scanning step: 50 mV, hold time: 1.5 s, delay time: 1.5 s to guarantee steady state. The devices were with 40 μm grating pitch and different gap distances. (a) Linear I-V curves showed larger current generated from smaller gap distances. (b) A voltage plateau around 0.9 V shown on the log I *vs.* V curves. (c) The plot of electrolysis current *vs.* gap distance$^{-1}$ at different voltages demonstrated that the pure water splitting was limited by electron-transfer due to the high electric field in the entire gap to enhance mass transport. (d) Bubble generation around 2 V. Sometimes bubble generation were very few, which may result from nanobubbles dissolved into water[42, 43].

In NECs the electron-transfer rate only depends on the cell voltage while the transport rate (mainly by migration) depends both on voltage and gap distance (*i.e.*, electric field). In a plot of electrolysis current *vs.* gap distance$^{-1}$ (a scale of electric field) at each voltage, if the reaction is limited by electron-transfer, the current should be relatively independent of the gap distance; however, if the reaction is limited by mass transport, the current should be sensitive to the gap distance (showing a large slope). Figure 4(c) clearly demonstrated such effects. For large gaps, a large slope appeared on the figure since the reaction was mass-transport limited; when the gap was small enough, the current reached saturation value only dependent on the voltage, indicating electron-transfer limited reaction. The critical gap distance (or "turning point") between such two states became smaller (moved to the right on the figure) with increasing voltage. This is because the electron-transfer rate increases faster than the mass transport when voltage increases (exponential *vs.* linear), therefore smaller gaps are necessary in order to achieve saturation current (electron-transfer limited) at higher voltages.

**Sodium hydroxide solution.** The electrolysis of pure water and 1 mol/L sodium hydroxide solution were compared in Figure 5(a), both based on our NECs with the same gap distance 72 nm and different grating pitches. For pure water, the electrolysis current at 1.8 V linearly increased with the number of grating edges; while for sodium hydroxide solution, the current was less dependent on the number of edges (*i.e.*, grating pitch) and the data dispersion was significantly larger than that of pure water. The mechanism is shown in Figure 5(b). For pure water splitting, the reaction only occurs at the edges where the electrode distance is small enough to couple the two half-reactions together; at the "non-edge" region (*i.e.*, top face) of the grating line, the scenario is just like pure water splitting in macrosystem (self-limited due to large electrode distance). On the contrary, in sodium hydroxide solution the entire surface was involved in supporting the reaction. That is because Debye-length in 1 mol/L sodium hydroxide solution is less than 1 nm, still significantly smaller than the electrode distance (72 nm here). Thus, the two half-reactions are not coupled together and are still diffusion-limited, just like that in macrosystem where the reactions occurs on all accessible parts of the electrodes. Therefore, the current greatly depends on the effective reaction area. In our present experiments, the area was that covered by the solution droplet and was not accurately controlled. Thus, the current from sodium hydroxide solution was not sensitive to grating pitches and presented significant variability (see more discussions in Figure S7). In comparison, the current from pure water was not sensitive to the area of the droplet region, and bubbles could only form within the grating region. Supplementary information also discusses the two voltage plateaus around 0.4 V and 1.2 V, respectively, on the log-plot of the I-V curves in sodium hydroxide solution.

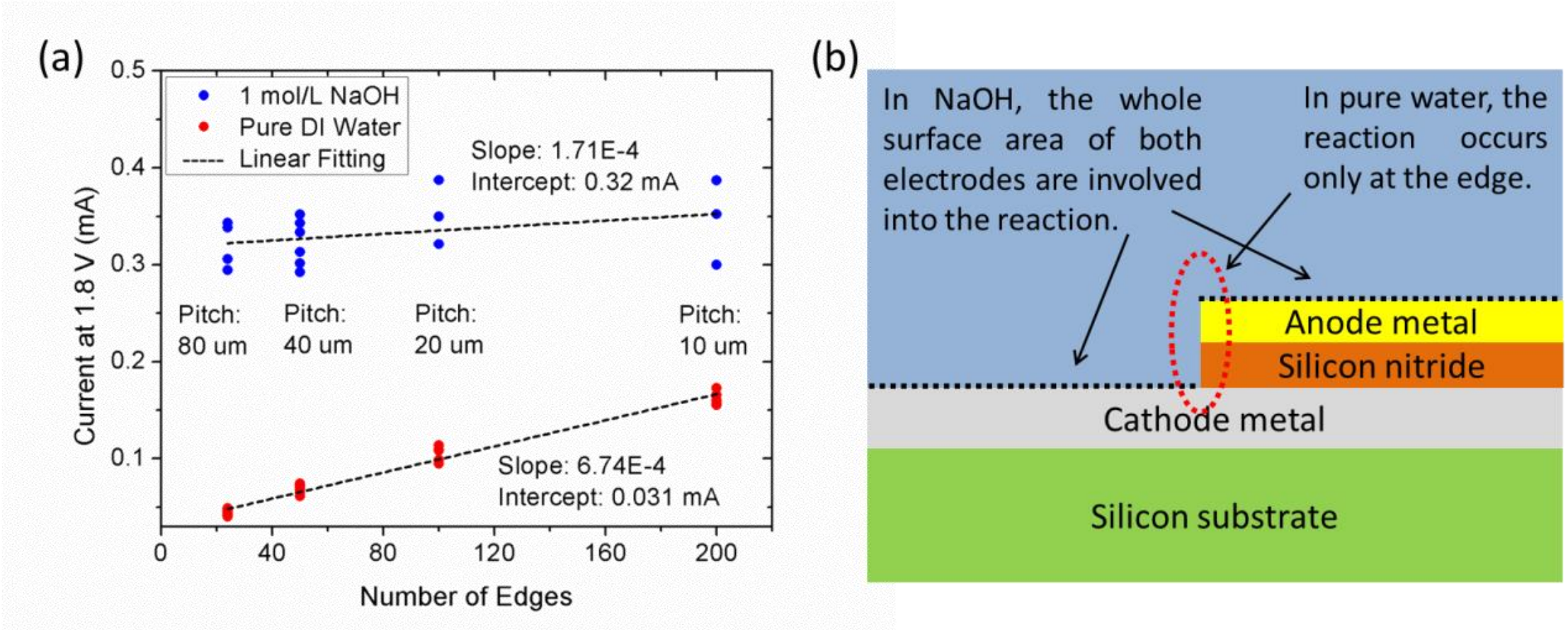

Figure 5. Comparison between pure water splitting and water splitting in 1 mol/L sodium hydroxide solution, both based on our NECs. The experiment conditions were the same in Figure 4. The devices were with 72 nm gap distance and different grating pitches. (a) The relationship between electrolysis current at 1.8 V and the number of edges. The number was calculated from the grating pitches since the grating region was fixed with width of 1 mm (each grating line had two edges). (b) Schematic diagram of the mechanisms of the different reaction locations for pure water splitting and water splitting in sodium hydroxide solution. In Figure 5(a), the slope (increased current per edge) from the pure water curve was almost 4 times of the slope from the sodium hydroxide solution curve, indicating much larger contribution to electrolysis current from field-driven effect than from diffusion effect. The extrapolated intercept value 0.32 mA of sodium hydroxide solution indicated the nature of entire surface involved into the reaction. For pure water, the background current 0.031 mA was much smaller, probably resulting from capacitive current or ionic impurities.

Notice that the current from sodium hydroxide solution still increased slightly with the number of edges. Even though it was diffusion limited, at the edges the overall diffusion length had shrunk to 72 nm. Therefore more edges could slightly enhance the electrolysis current. Notice that the effective reaction area in sodium hydroxide solution was much larger than that in pure water. Even under such unfavorable conditions, the electrolysis current from pure water was comparable to that from 1 mol/L sodium hydroxide solution, indicating more than $10^5$-fold enhancement of the apparent conductivity of pure water. (The conductivity of 1 mol/L sodium hydroxide solution and common pure water (equilibrated with $CO_2$ in air) are $2\times10^5$ μS/cm and 1 μS/cm[39], respectively.) From the linear fitting, we can conclude that when the grating pitches are smaller than 2 μm, the electrolysis current from pure water can be even higher than that from 1 mol/L sodium hydroxide solution of the same pitch (2 μm pitch is beyond our photolithography capability, therefore we did not attempt it at present stage). These results demonstrate the enhancement of pure water splitting by virtual breakdown effect compared to conventional transport-limited reaction and a potential for greatly increased efficiency for hydrogen production.

## Conclusion

Field-driven pure water splitting at room temperature has been successfully achieved in this paper based on our metal-dielectric-metal sandwiched-like nanogap electrochemical cells. The gap distance between anode and cathode down to 37 nm has been demonstrated. In such deep-sub-Debye-length region, high electric field in the entire gap significantly enhances water molecules ionization and mass transport, leading to an electron-transfer limited reaction. This virtual breakdown mechanism can greatly enhance the equivalent conductivity of pure water to more than $10^5$-fold, resulting in electrolysis current comparable to or even high than that from 1 mol/L sodium hydroxide solution, and thus a higher efficiency for hydrogen production. We propose to investigate this virtual breakdown mechanism further. For example, reference electrode can be added to study cathode current and anode current separately; characterizations of capacitance-voltage curves will also provide important information for theoretical analysis. Moreover, such virtual breakdown mechanism can be applied on almost all weakly-ionized materials, and may have applications for ultrafast charging, alcohol electrolysis, carbon-dioxide reduction and fuel cells. Besides, compared to other NECs, our open cells can be simply fabricated on large area with high yield, and have great potentials to enhance the rate of redox reactions for ultra-sensitivity/selectivity. At last, compared to current industrial water electrolysis, such high-efficiency pure water splitting without any electrolyte at room temperature, especially connected to renewable energy sources, is very promising for both mass manufacturing and portable devices for on-demand clean hydrogen production.

## Acknowledgements

The authors thank Prof. Stephen B. Cronin, Prof. Rehan Kapadia and Prof. Mark Thompson at the University of Southern California for insightful discussions. The authors also thank Prof. Chongwu Zhou at the University of Southern California for use of the probe station.

## Author contributions

W.W. led the project. Y.W. and W.W. conceived the research and put forward the theoretical model. Y.W. and S.R.N. designed the experiments. Y.W. prepared the devices and performed the characterization. All the authors contributed to analyzing and interpreting the data, and to writing the manuscript.

## Additional information

Supplementary information is provided.

## Competing financial interests

The authors declare no competing financial interests.

# Supplementary information

## 1. Finite element calculations

The simulation results shown in Figure 1 were achieved using commercial software, Comsol Multiphysics® 5.2. The 2-D geometry and boundary conditions setting are shown in Figure S1 (take gap distance of 5 μm as an example, only one boundary edge of our sandwiched-like nanogap cells was simulated).

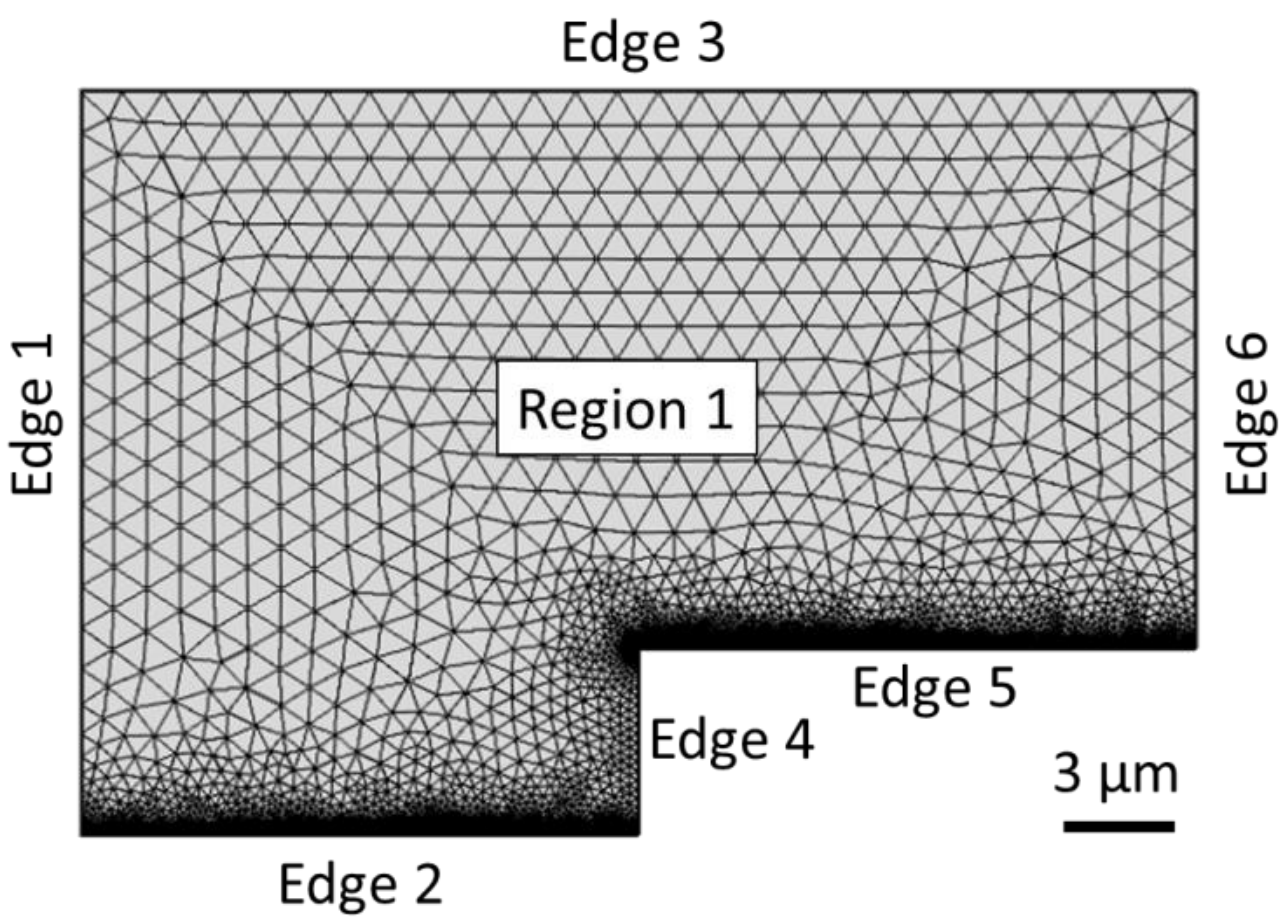

| Boundary | Electrostatics | Species Transport |
|---|---|---|
| Edge 1 | Zero charge | No flux |
| Edge 2 | rho_surf, deltaphi, phiM ( at cathode) | No flux |
| Edge 3 | Ground | c_H_bulk, c_OH_bulk |
| Edge 4 | Zero charge | No flux |
| Edge 5 | rho_surf, deltaphi, phiM (at anode) | No flux |
| Edge 6 | Zero charge | No flux |
| Region 1 | rho_space | c_H_bulk, c_OH_bulk (initially) |

Figure S1. Geometry and boundary conditions setting in finite element calculations.

The parameters setting are shown in Table I.

Table I. Parameters setting in finite element calculations.

| Name | Value | Unit | Description |
|---|---|---|---|
| T0 | 25 | degC | Temperature |
| c_H_bulk | 0.0001 | mol/m^3 | Bulk cation concentration |
| c_OH_bulk | c_H_bulk | | Bulk anion concentration |
| z_H | 1 | | Cation charge |
| z_OH | -1 | | Anion charge |
| D_H | 9.31E-09 | m^2/s | Diffusion coefficient, cation |
| D_OH | 5.26E-09 | m^2/s | Diffusion coefficient, anion |
| eps_H2O | 80 | | Relative permittivity of water |
| xS | 0.2 | nm | Stern layer thickness |
| phi_anode | 0.5 | V | Anode potential |
| rho_space | F_const*(z_H*c_H+z_OH*c_OH) | C/m³ | Space charge density |
| deltaphi | phiM-phi | V | Electrode-OHP potential difference |
| rho_surf | epsilon0_const*eps_H2O*deltaphi/xS | C/m² | Surface charge density |
| phiM (at anode) | phi_anode/2 | V | Anode potential |
| phiM (at cathode) | -phi_anode/2 | V | Cathode potential |
| thk_nitride | Manually setting | μm | Thickness of silicon nitride layer |

The equations that governed the ions movement and distribution were the steady-state Nernst-Planck equation and the Poisson equation,

$$J_i = -D_i \nabla C_i - (z_i F / RT) D_i C_i \nabla \varphi \tag{S1}$$

$$\nabla^2 \varphi = -\rho / \varepsilon_r \varepsilon_0 \tag{S2}$$

where $J_i$, $D_i$, $C_i$, and $z_i$ are the current density, diffusion coefficient, concentration and charge of species *i*, $\varphi$ is the local electric potential, $\rho$ is the local net charge density in the solution, $\varepsilon_r$ is the static dielectric constant, $\varepsilon_0$, *F*, *R*, and *T* are the permittivity of vacuum, Faraday constant, gas constant and temperature. To simplify the problem, $\varepsilon_r$ of pure water was set constant 80 in the entire solution even though near the electrode surface $\varepsilon_r$ can be reduced to less than 10[1].

The Debye-length (around 1 μm in pure water) was calculated from Gouy-Chapman theory, which requires infinite electrode plane and potential much smaller than 26 mV at room temperature. Simulation results showed that, even though our modeling could not satisfy the two requirements of Gouy-Chapman theory, the approximation value of 1 μm could still be used since little difference showed up between the theoretical value (from the Gouy-Chapman theory) and simulated value (from the software simulation). Besides, the smallest nanogap between the two electrodes we achieved was 37 nm, which was much smaller than both theoretical value and simulated value. Thus, the claim of "deep-sub-Debye-length" is still valid.

Stern layer had been considered in the initial setting; however, the final results had little dependence on with or without Stern layer setting. This is probably because the simulation mesh was not fine enough near the electrode surface. Mesh quality is a key factor of the simulation results. We discovered that finer mesh near the surface greatly enhanced the surface concentration (more obvious when large potential added). However, further finer meshing was not possible due to limited computational resources. Here, more accurate results might not be necessary. Quantitatively, we have demonstrated the double layer overlapping effect, and high electric field (just voltage divided by gap distance) uniformly distributed in the entire gap has been demonstrated as well. For our current research, we determined that these simulation results are sufficient.

## 2. Virtual breakdown effect

For pure water splitting in nanogap cells, the two half-reactions are coupled together. Take the anode as an example. At the anode $OH^-$ ions (the reaction ions) come from two parts: one is from water ionization near the anode; the other part comes from the $OH^-$ ions migrated from the cathode to the anode. When the gap is small enough, migration rate can be larger than electron-transfer rate so that the reaction is limited by electron-transfer. Initially, 1 unit of electron-transfer leads to two units of $OH^-$ ions generation, resulting in non-consumed $OH^-$ ions accumulated near the anode. Such ions accumulation at the electrodes slows down further water ionization near the electrodes, to reduce the total $OH^-$ ions generation rate to balance with the 1 unit of electron-transfer in the external circuit. Under steady state condition, the sum of the $OH^-$ ions from the two parts is balanced with the 1 unit of electron-transfer in the external circuit. Such scenario appears like that the water molecules are split into $H_3O^+$ and $OH^-$ ions in the middle of the gap, allowing $H_3O^+$ ions to drift towards the cathode and $OH^-$ ions to drift towards the anode, respectively. The whole effect looks like that water has been broken-down. However, we should point out that in fact water molecule dissociation still occurs only near the electrode (due to local ion consumption); the water dissociation in the middle of the gap is just the equivalent effect.

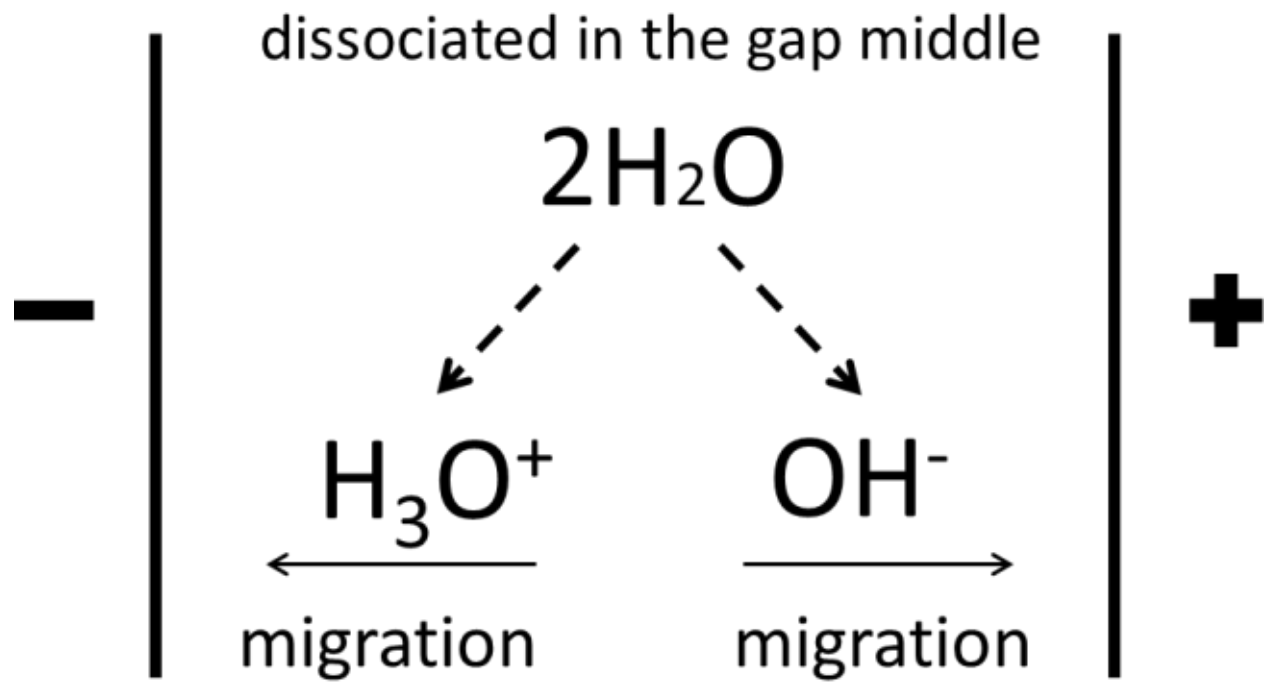

Figure S2. The equivalent effect that water molecules dissociated in the middle of the gap.

## 3. RC-circuit model

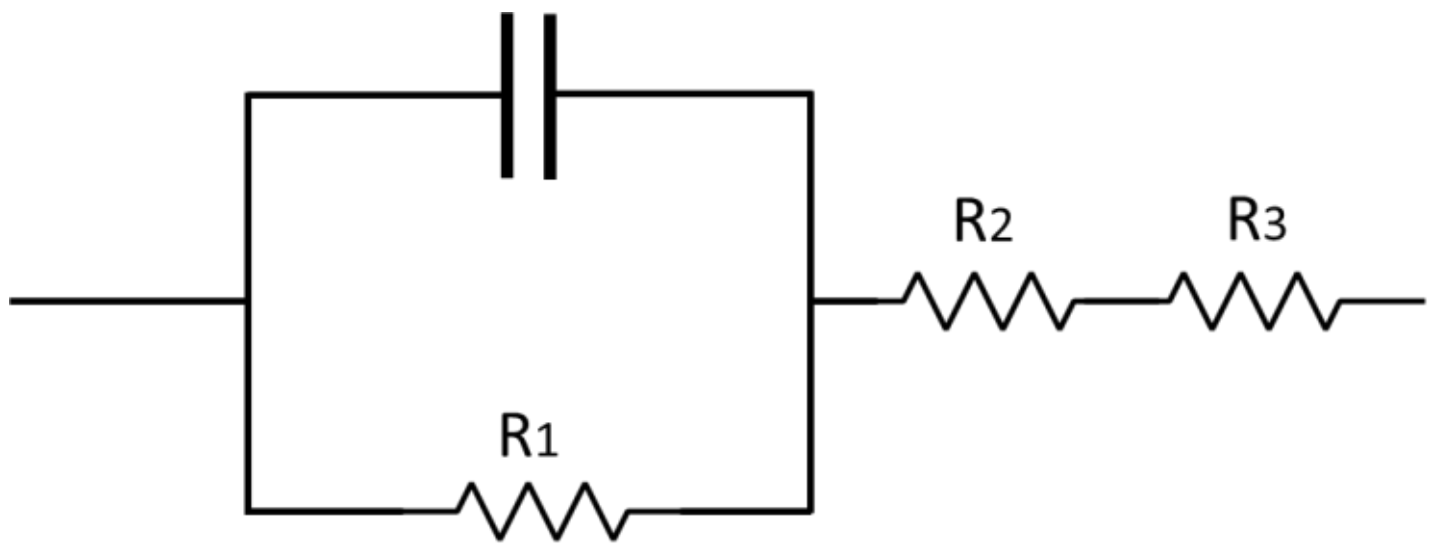

Figure S3. RC-circuit model of half-reaction of water splitting in nanogap cells.

Figure S3 shows the RC-circuit model of half-reaction of water splitting in nanogap cells. The capacitor represents the double layer. $R_1$ represents the reaction rate of electron-transfer, which depends only on the changes in potential drop across the interface. $R_2$ represents the mass transport rate, which is related to both voltage and gap distance (*i.e.*, electric field in the gap). $R_3$ represents the water ionization rate. When gap distance is smaller, $R_2$ becomes smaller; while $R_1$ can also be slightly smaller since local reaction ion concentrations become higher and potential drop becomes larger, meaning that $R_1$ is coupled to $R_2$. $R_3$ may depend on $R_1$ because ions are consumed continuously, $R_3$ can be enhanced by shifting the ionization equilibrium. When the gap distance is around Debye-length, $R_2$ is the largest one and determines the whole reaction rate; however, when gap is much smaller than Debye-length, $R_2$ can be smaller than $R_1$, indicating electron-transfer limited reaction. That is to say, when gap distance further decreases, the current reaches a saturation value that only depends on voltage.

## 4. Low DC-bias silicon nitride anisotropic etching

To avoid short-circuit between top anode and bottom cathode metal layers, low

DC-bias silicon nitride etching technology was developed in order to reduce the ion bombardment effect. In experiments we discovered that traditional nitride etching with high DC-bias could lead to low yield of device fabrication: most of the devices got short-circuit after nitride etching. This was because that the sputtered metal atoms formed short-circuit path on the sidewall, connecting top anode and bottom cathode[2]. By using our low DC-bias etching recipe, the fabrication yield has been improved greatly. The recipe parameters and the etching profile are shown in Figure S4. The DC-bias of the silicon nitride etching was down to 19-21 V with etching rate larger than 80 nm/min and vertical sidewalls, even better than our previously published[2]. Exactly vertical sidewalls were not required because in fact a little bit isotropic etching was desired since anode tips at the boundary could form higher electric field.

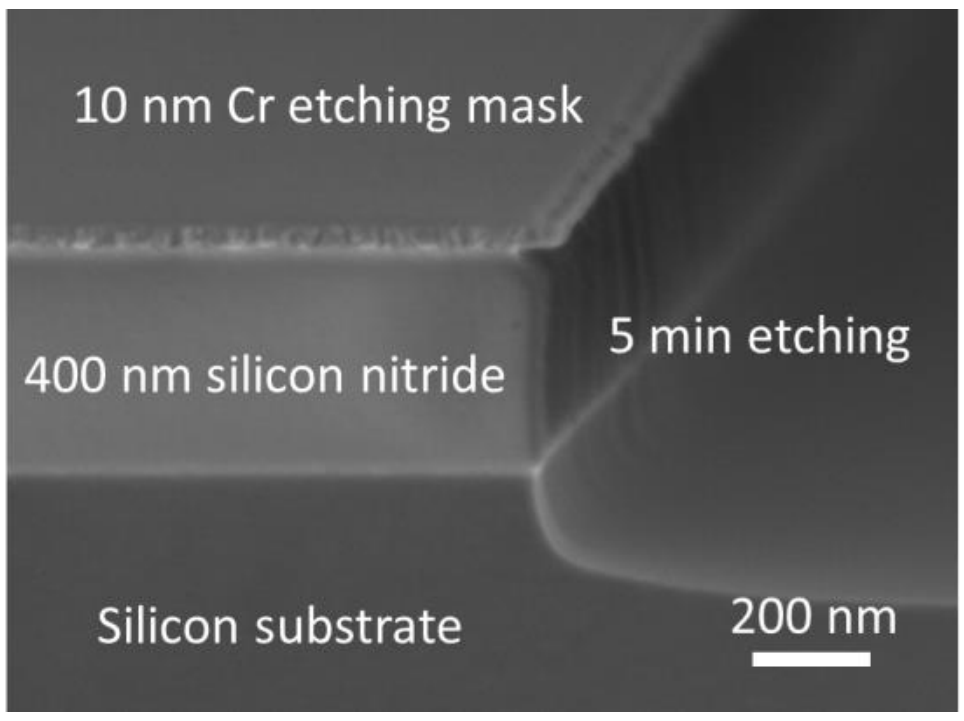

| | | | |
|---|---|---|---|
| $SF_6$ (sccm) | 8.0 | $C_4F_8$ (sccm) | 10.0 |
| Pressure (mT) | 7.0 | Temp (° C) | 20 |
| Strike RF (W) | 40 | Strike ICP (W) | 2300 |
| Strike Time (s) | 5 | Etch RF (W) | 6 |
| Etch ICP (W) | 2300 | DC Bias (V) | 19-21 |
| Etching Rate | > 80 nm/min | Vertical Sidewalls | |

Figure S4. The recipe parameters and the etching profile of our low DC-bias silicon nitride etching.

In experiments, wet etching (without Cr mask) of silicon nitride using buffered oxide etch (7:1) was attempted as well because it naturally avoids the ion bombardment effect. However, the two electrodes always got short-circuit by using this method, which may be attributed to capillary contact of the two electrodes because of lateral etching undercut. Therefore this method was given up.

## 5. Anode damage

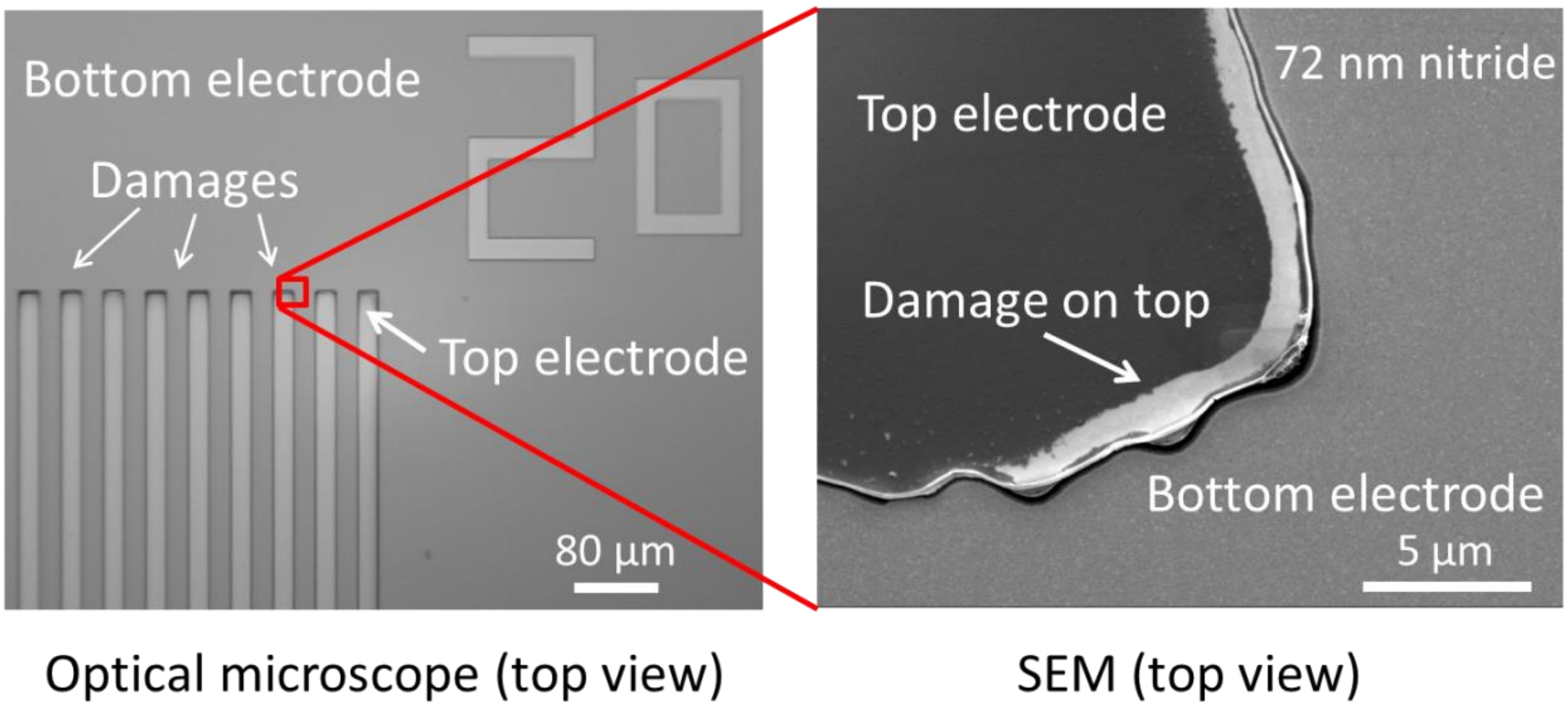

Figure S5. Anode damage in pure water when voltage above 5 V. The device shown here is with 72 nm gap and 40 μm pitch.

Anode can be roughed during redox cycling[3, 4]. The roughness comes from the electrochemically oxidation and re-deposition of the anode metal, even for gold[3, 5]. In experiments, such anode damage sometimes occurred when the applied voltage values were above 5 V (Figure S5). Thinner-gap samples were more likely to suffer damage. Moreover, damage always showed up near the grating boundary where the electric field was the highest. Such damage, especially the re-deposition of gold atoms, can lead to short-circuit between the anode and the cathode (especially for smaller gap distances), and thus reducing the lifetime of the devices. To avoid such short-circuit, the maximum external voltage was set to be 2.5 V (to reduce the current density in fact). In this way, the devices can be measured repetitively without obvious damage or short-circuit.

Two possible approaches are proposed here to avoid such anode damage. First, indium tin oxide (ITO) can be used to replace the gold as anode material, with its highest oxidation state which cannot be oxidized further[6, 7]. Second, an ultrathin layer of energy-band offset material[8, 9] may be coated onto the gold anode, with thickness small enough to be conductive to gold while preventing gold contacting water directly.

## 6. Bubble effects

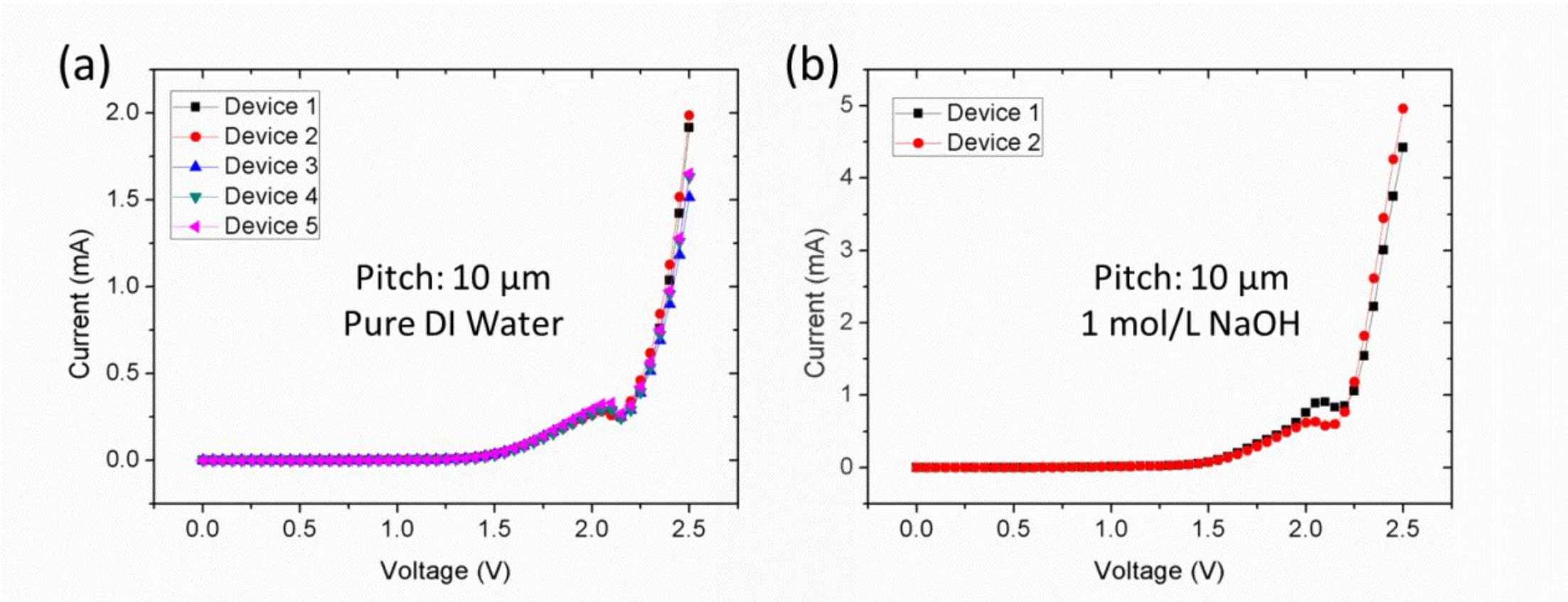

Figure S6. Bubble effects on plateaus (or peaks) around 2 V in I-V curves based on devices with 72 nm gap and 10 μm pitch, (a) in pure water, (b) in sodium hydroxide solution.

Figure S6 shows plateaus (or peaks) around 2 V in I-V curves, both in pure water measurements and sodium hydroxide solution measurements. We believe that it was due to bubble effects. Around 2 V, bubble generation started to be vigorous enough so that it could be observed by the naked eye. Moreover, devices with smaller gap distance or smaller grating pitches could have more obvious plateaus around 2 V, indicating that such plateaus were determined by the geometry of the structures, rather than electrode electrochemical reactions. This observation is reasonable to expect since bubbles are more likely to be trapped within the smaller gap or smaller pitch structures before releasing, excluding the water involved in the reaction. Therefore, larger voltage leads to larger excluding effect, and thus smaller current, showing negative resistance which performs like a plateau or peak in I-V curves.

Notice that Figure S6(a) also shows the consistent results among several tests. The data from different devices were almost exactly the same, especially below 2 V. Above 2 V, the data had a relatively larger error range which we think was due to the bubble effects on current performance. Therefore, we always selected the current data below 2V for analysis and comparison to be free from bubble related artifacts.

## 7. Sodium hydroxide solution: reactions in the entire surface

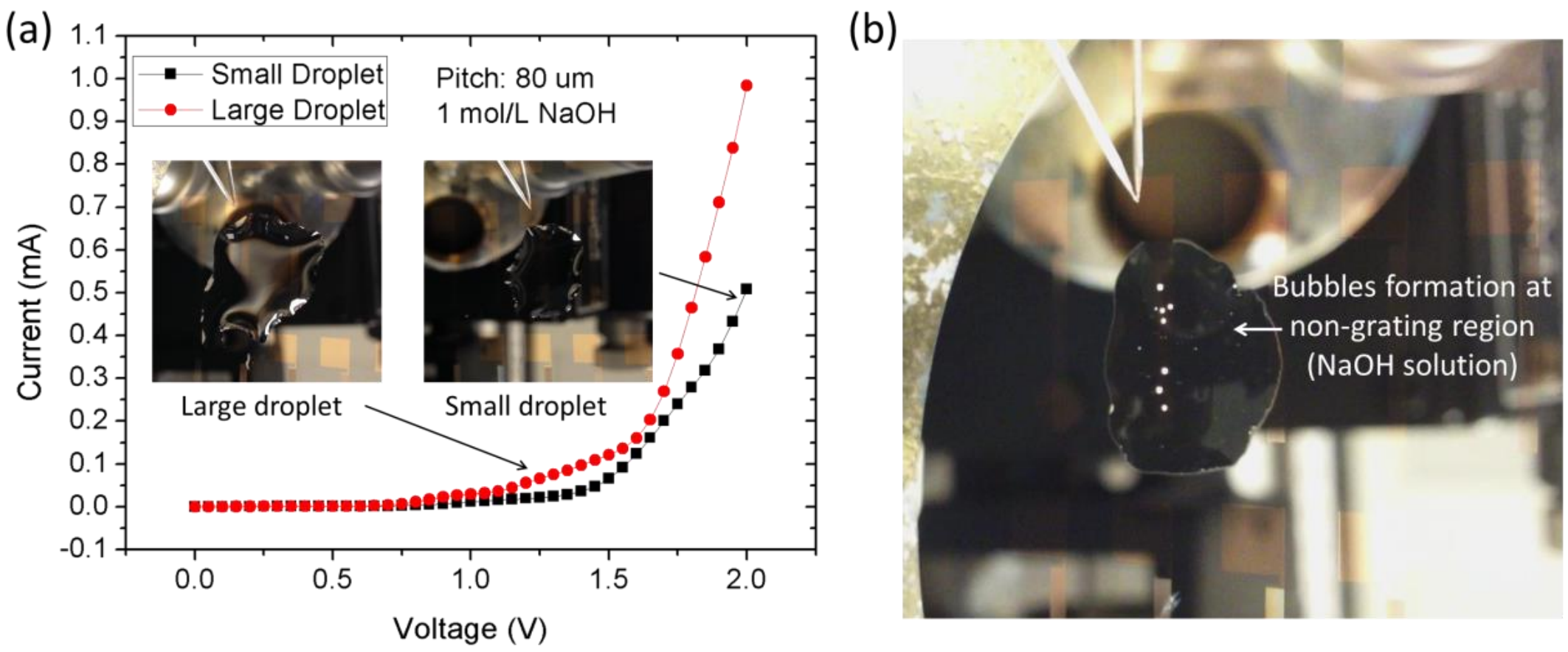

Figure S7. Evidence of the entire surface involved into the reactions in sodium hydroxide solutions. (a) Large droplet provided large current. (b) Bubbles formed at non-grating region. The devices were with 72 nm gap distance.

Reactions in pure water only occur at the edge boundary of each grating line in our sandwiched-like nanogap cells. Different from pure water, the entire surface is involved in the reactions in sodium hydroxide solutions. Two types of evidence are shown in Figure S7. First, the larger droplet of sodium hydroxide solution provided larger current, indicating more surface area involved into the reactions, though the total number of the edges was independent of the droplet size. Second, bubbles could be even generated far away from counter-electrode (*i.e.*, non-grating region), demonstrating that reactions can occur even very far from the grating edges. That is to say, the reactions in sodium hydroxide solutions not only occur at the grating edges, but also over the entire region covered by the droplet.

## 8. Plateaus in log I *vs.* V curves

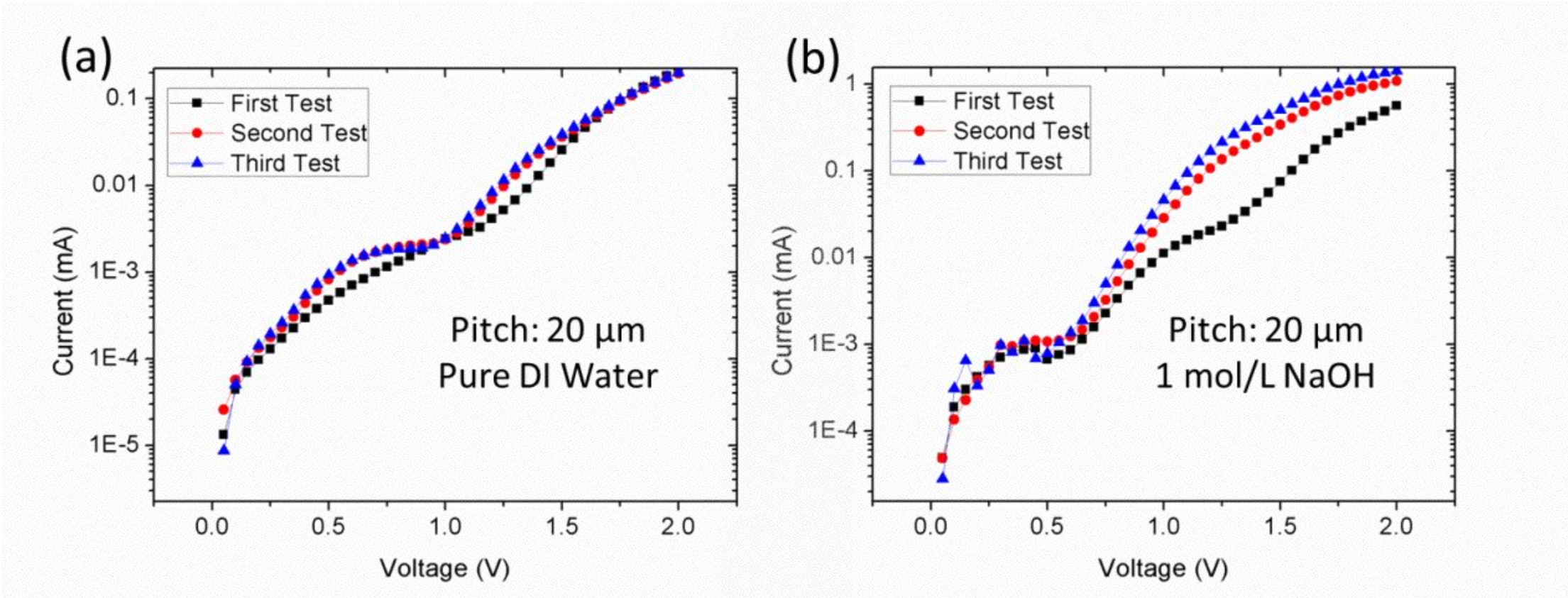

Figure S8. Plateaus in log I *vs.* V curves from (a) pure water tests and (b) sodium hydroxide solution tests. The devices were with 72 nm gap distance.

Figure S8 shows the log I *vs.* V curves from tests of both pure water and sodium hydroxide solution. For pure water, one plateau appeared around 0.9 V (also shown in Figure 4(b)). This plateau became flatter after the first test on the same device. For sodium hydroxide solution, two plateaus, around 0.4 V and 1.2 V respectively, were shown on the log I *vs.* V curves. After first test, the 0.4 V plateau still existed but the 1.2 V plateau disappeared, and the current became much larger (the 2V plateau can be ignored since it is due to the bubble effects). This plateau phenomenon is quite repeatable, no matter what the gap distance or pitch is, indicating that it is more likely related to the intrinsic electrochemical reactions, rather than geometry factors. However, the fundamental mechanism is not clear.

Only a few literature reviews discussed about such plateaus in log I *vs.* V curves. Our hypothesis is the following. The 0.9 V plateau from pure water tests may be attributed to dissolved oxygen reduction or anode gold oxidation (and these two effects might be coupled). For sodium hydroxide solution, the 0.4 V plateau most likely came from the reduction of dissolved oxygen; while the 1.2 V plateau was related to anode gold oxidation. The different values of the oxidation plateaus in pure water and in sodium hydroxide solution was most likely due to the difference in pH values. The 0.4 V plateau would not disappear since for every test new sodium hydroxide solution (without inert gas saturation) was used. For the gold anode, non-conductive oxide state I and conductive oxide state II can form during water splitting[10]. During the first test in sodium hydroxide solution, $OH^-$ ions concentration was so large that all surface gold could be oxidized to state II, therefore during the second or third tests no surface gold could be oxidized further (thus the 1.2 V plateau disappeared). Also, because oxide state II was porous and conductive, the distance between anode and cathode had been shortened due to gold oxide islands and the current after the first test could become larger (the larger current could be also attributed to roughness of the surface since effective reaction area became larger[4]). However, for pure water, $OH^-$ ions concentration was small so that only oxide state I might form, therefore gold could still be oxidized further into the formation of state I during the second or even third tests (until two or three monolayers of the oxide state I coverage reached[3, 11]), with almost the same electrolysis current or smaller since oxide state I was non-conductive. However, such plateaus may also result from the formation of oxygen coverage[12], inhibition layer[13] or inert sites[14].

Detailed experiments are necessary to get a clearer fundamental understanding of the mechanism underlying such plateaus. First, inert gas saturated pure water and sodium hydroxide solution should be utilized; second, anode current and cathode current should be studied separately; third, crystal plane of original gold and final anode oxidation should be further analyzed by spectroscopy measurement. However, since this problem is beyond the scope of our present study, we have not included such experiments in this paper.